\documentclass[a4paper,12pt]{article}
\usepackage{graphicx,rotating,amsmath,amssymb}
\usepackage{hyperref}\usepackage{slashed,cite}
\pdfoutput=1

\hypersetup{colorlinks,bookmarksopen,bookmarksnumbered,
linkcolor=blus,pdfstartview=FitH,urlcolor=rossos,citecolor=verde}

\newcommand{\fig}[1]{~\ref{fig:#1}}

\usepackage{multicol}
\usepackage{color}
\definecolor{rosso}{cmyk}{0,1,1,0.4}
\definecolor{rossos}{cmyk}{0,1,1,0.55}
\definecolor{rossoc}{cmyk}{0,1,1,0.2}
\definecolor{blu}{cmyk}{1,1,0,0.3}
\definecolor{blus}{cmyk}{1,1,0,0.6}
\definecolor{bluc}{cmyk}{1,1,0,0.1}
\definecolor{verde}{cmyk}{0.92,0,0.59,0.25}
\definecolor{verdec}{cmyk}{0.92,0,0.59,0.15}
\definecolor{verdes}{cmyk}{0.92,0,0.59,0.4}
\newcommand{\eq}[1]{~{\rm (\ref{eq:#1})}}

\newcommand{\MeV}{\,{\rm MeV}}

\newcommand{\TeV}{\,{\rm TeV}}

\def\circa#1{\,\raise.3ex\hbox{$#1$\kern-.75em\lower1ex\hbox{$\sim$}}\,}

\newcommand{\beq}{\begin{equation}}
\newcommand{\eeq}{\end{equation}}

\newcommand{\bea}{\begin{eqnarray}}
\newcommand{\eea}{\end{eqnarray}}
\newcommand{\be}{\begin{equation}}
\newcommand{\ee}{\end{equation}}
\font\tenrsfs=rsfs10 at 12pt
\font\sevenrsfs=rsfs7
\font\fiversfs=rsfs5
\newfam\rsfsfam
\textfont\rsfsfam=\tenrsfs
\scriptfont\rsfsfam=\sevenrsfs
\scriptscriptfont\rsfsfam=\fiversfs
\def\mathscr#1{{\fam\rsfsfam\relax#1}}

\def\circa#1{\,\raise.3ex\hbox{$#1$\kern-.75em\lower1ex\hbox{$\sim$}}\,}
\makeatletter

\def\hhref#1{\href{http://arxiv.org/abs/#1}{arXiv:#1}} 
\def\baselinestretch{1.11}
\def\hhref#1{\href{http://arxiv.org/abs/#1}{arXiv:#1}}
\usepackage{xstring}
\newcommand{\hhrefq}[1]{\IfSubStr{#1}{:}{\href{http://inspirehep.net/search?ln=en&ln=en&p=#1&of=hb&action_search=Search&sf=&so=d&rm=&rg=25&sc=0}{InSpire:#1}}{\hhref{#1}}}

\def\art{\@ifnextchar[{\eart}{\oart}}
\def\eart[#1]#2#3#4#5#6{{\rm #2}, {\em #3 \bf #4} {\rm (#6) #5} ({\em #1})}
\def\article{\@ifnextchar[{\earticle}{\oarticle}}
\def\oarticle#1#2#3#4#5#6{{\rm #1}, {\ital `#6'}, {\rm #2 #3 (#5) #4}}
\def\earticle[#1]#2#3#4#5#6#7{{\rm #2}, {\ital `#7'}, {\rm #3 #4 (#6) #5}  [\hhrefq{#1}]}
\def\hepart[#1]#2{{\rm #2, \sl#1}}
\def\heparticle[#1]#2#3{#2, {\ital `#3'} [\hhrefq{#1}]}
\newcommand{\doi}[1]{\href{http://dx.doi.org/#1}{[link]}}
\newcommand{\hhrefqq}[1]{\IfBeginWith{#1}{10.}{\href{https://doi.org/#1}{doi:#1}}{\hhrefq{#1}}}

\renewenvironment{thebibliography}[1]
{\begin{multicols}{2}[\section*{\refname}]%
		\@mkboth{\MakeUppercase\refname}{\MakeUppercase\refname}%
		\list{\@biblabel{\@arabic\c@enumiv}}%
		{\settowidth\labelwidth{\@biblabel{#1}}%
			\leftmargin\labelwidth
			\advance\leftmargin\labelsep
			\@openbib@code
			\usecounter{enumiv}%
			\let\p@enumiv\@empty
			\renewcommand\theenumiv{\@arabic\c@enumiv}}%
		\sloppy
		\clubpenalty4000
		\@clubpenalty \clubpenalty
		\widowpenalty4000%
		\sfcode`\.\@m}
	{\renewcommand{\@noitemerr}
		{\@latex@warning{Empty `thebibliography' environment}}%
		\endlist\end{multicols}}

\font\ital=cmu10

\newcounter{alphaequation}[equation]
\def\thealphaequation{\theequation\hbox to
0.6em{\hfil\alph{alphaequation}\hfil}}
\def\eqnsystem#1{
\def\@eqnnum{{\rm (\thealphaequation)}}
\def\@@eqncr{\let\@tempa\relax \ifcase\@eqcnt \def\@tempa{& & &} \or
  \def\@tempa{& &}\or \def\@tempa{&}\fi\@tempa
  \if@eqnsw\@eqnnum\refstepcounter{alphaequation}\fi
\global\@eqnswtrue\global\@eqcnt=0\cr}
\refstepcounter{equation} \let\@currentlabel\theequation \def\@tempb{#1}
\ifx\@tempb\empty\else\label{#1}\fi
\refstepcounter{alphaequation}
\let\@currentlabel\thealphaequation
\global\@eqnswtrue\global\@eqcnt=0 \tabskip\@centering\let\\=\@eqncr
$$\halign to \displaywidth\bgroup \@eqnsel\hskip\@centering
$\displaystyle\tabskip\z@{##}$&\global\@eqcnt\@ne
\hskip2\arraycolsep\hfil${##}$\hfil& \global\@eqcnt\tw@\hskip2\arraycolsep
$\displaystyle\tabskip\z@{##}$\hfil
\tabskip\@centering&\llap{##}\tabskip\z@\cr}
\def\endeqnsystem{\@@eqncr\egroup$$\global\@ignoretrue} \makeatother

\newcommand{\SU}{\,{\rm SU}}

\begin{document}
\begingroup
\renewcommand{\baselinestretch}{1}
\begin{center}
{\LARGE \bf \color{rossos}
QCD corrections to Minimal\\[1ex] Dark Matter annihilations}\\
\bigskip\bigskip
{\large\bf Alessandro Strumia}
\\[2ex]
{\it Dipartimento di Fisica dell'Universit{\`a} di Pisa, Italia}\\

\bigskip\bigskip

{\large\bf\color{blus} Abstract}
\begin{quote}\large
QCD corrections to fermionic Minimal Dark Matter annihilations increase its annihilation cross section
(by 2\% for a weak doublet, 1.3\% for a  triplet, 0.5\% for a quintuplet) and thereby the
Dark Matter mass required to achieve the observed cosmological relic abundance via thermal freeze-out.
\end{quote}
\thispagestyle{empty}
\end{center}
\endgroup
\setcounter{footnote}{0}

\subsubsection*{}
Adding to the Standard Model  one fermionic electro-weak $n$-plet
provides a motivated and predictive Minimal theory of Dark Matter (MDM), characterized by a single free parameter: the dark matter mass $M$~\cite{hep-ph/0512090,2406.01705}.
A quintuplet is accidentally stable on cosmological timescales. 
A triplet with hyper-charge
$Y=0$ or a doublet with $Y=1/2$ can be rendered stable by imposing a $\mathbb{Z}_2$ symmetry,
and are further motivated in supersymmetric contexts as the ‘wino’ or `higgsino' (see e.g.~\cite{[hep-th/0405159}).

In each case the dark matter mass is unambiguously determined by requiring that the observed relic abundance $\Omega_{\rm DM} h^2 = 0.120\pm0.001$~\cite{2406.01705}
originates from thermal freeze-out in standard cosmology. 
For $n\ge 3$ order unity corrections arise from electro-weak Sommerfeld enhancements~\cite{hep-ph/0610249,0706.4071}  and related bound states~\cite{1702.01141}. 

A more precise determination of the MDM mass predicted by cosmology is motivated by: 
i) the  predictivity of the MDM framework;
ii)  the DM production rate at a muon collider is enhanced if operated at the specific energy 
that allows the resonant production of  DM-DM bound states~\cite{2103.12766};
iii)  indirect DM detection rates strongly depend on $M$ in view of the following numerical coincidence.
For the triplet and quintuplet, the DM mass is predicted to be 
$M \approx 2.8\TeV$ and $M\approx 13.7\TeV$ respectively, see table~\ref{tab:M235}.
Sommerfeld and bound state corrections enhance indirect detection signals  by orders of magnitude if
$M \approx 2.5\TeV$ for the triplet and $M\approx 13.4\TeV$ for the quintuplet~\cite{1702.01141,2507.17607}.
As a result of this proximity, the lower side of the allowed quintuplet thermal mass range is disfavoured by indirect detection data~\cite{1702.01141,2507.17607}.\footnote{Safe bounds
are obtained by requiring  that the DM indirect detection signals, computed conservatively assuming  cored DM profiles, do not exceed the total observed fluxes~\cite{0706.4071,1702.01141}.
Recent analyses~\cite{2507.15934,2507.17607} attempt to improve the sensitivity by subtracting, from the measured fluxes, 
estimated astrophysical backgrounds with uncertain spectral and morphological properties.
After these subtractions, according to~\cite{2507.15934}, the  thermal quintuplet mass range at $1\sigma$ is excluded for certain DM profiles.
However this claim relies on negative subtracted fluxes and (according to~\cite{2507.17607}) on over-estimated signal rates.}

\begin{table}[t]
\begin{center}
$$\small\begin{array}{c|ccc}
\hbox{DM mass}& \hbox{fermion doublet, $Y=1/2$} & \hbox{fermion triplet, $Y=0$} & \hbox{fermion quintiplet, $Y=0$}\\ \cline{2-4}
\hbox{predicted by}& 1.05\TeV\hbox{\cite{0706.4071}} & 2.8\TeV~ \hbox{\cite{1702.01141}} & 14\TeV~\hbox{\cite{1702.01141}} \cr
\hbox{cosmology} & (1.08\pm 0.02)\TeV\hbox{\cite{2205.04486,2410.02723}} &(2.86\pm 0.04) \TeV~\hbox{\cite{2107.09688,2410.02723}}  & (13.7^{+0.6}_{-0.3}) \TeV~\hbox{\cite{2107.09688,2410.02723}}\\
\hline
\hbox{QCD correction} & +1\% & +0.6\% & +0.25\%
\end{array}$$
\caption{\em\label{tab:M235} Minimal DM masses such that the relic DM abundance matches the observed DM cosmological abundance.
The computations in the second row include NLO corrections to potentials and an estimate of the $\pm 1\sigma$ thermal mass range.
The bottom row shows the QCD correction computed in this paper. 
}
\end{center}
\end{table}

QCD corrections to DM annihilations have been computed in supersymmetric models with various free parameters~\cite{hep-ph/0605181,hep-ph/0608215,0710.1821}.
We here compute QCD corrections to Minimal DM annihilations.
The Minimal Dark Matter (co-)annihilation cross section receives a contribution from final-state $q\bar{q}$ quarks via tree-level
$s$-channel exchange of electroweak vectors.
For fermionic dark matter, this channel contributes to the $s$-wave annihilation rate that dominates in the relevant non-relativistic limit.
In contrast, if dark matter is scalar, annihilations into quarks are $p$-wave suppressed and remain so even after including QCD corrections.

\begin{figure}[t]
$$\includegraphics[width=0.9\textwidth]{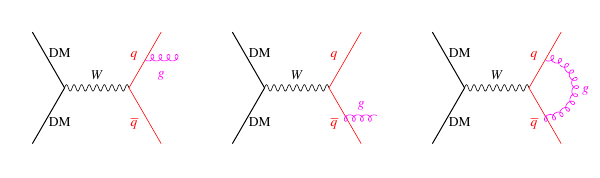}$$
\begin{center}
\caption{\em\label{fig:MDMQCDannihilation}Feynman diagrams for QCD corrections to Minimal DM annihilations.}
\end{center}
\end{figure}

The one-loop QCD correction consists of the sum of real gluon emission plus virtual corrections, as illustrated in fig.\fig{MDMQCDannihilation}.
As is well known, both contributions individually contain infra-red divergences ---
or, equivalently, terms enhanced by powers of $\ln(M/m_q)$ --- that cancel in the total rate.
Accordingly, we perform the computation in the limit of massless quarks $m_q=0$, 
using the Feynman gauge and dimensional regularisation in $d=4-2\epsilon$ dimensions
with $\overline{\rm MS}$ scale $\bar\mu$.

In the non-relativistic $s$-wave limit, the rate for the process
\beq {\rm DM}(p_1)\,\overline{\rm DM}(p_2) \to q(q_1) \bar q(q_2) g(q_3)\eeq 
 depends only on the total incoming momentum $P = p_1+p_2=q_1+q_2+q_3$ 
 with $P^2 \simeq 4 M^2$, rather than on the individual momenta $p_1$ and $p_2$.
We integrate the rate over the 3-body phase space
\beq \label{eq:dPhi3x1x2}
d\Phi_3 = \frac{M^2 e^{\gamma_E(4-d)}}{16(2\pi)^3} \left(\frac{M^2}{\bar\mu^2}\right)^{d-4} \frac{\left[
(1-x_1)(1-x_2)(1-x_3)\right]^{d/2-2}}{\Gamma(d-2)} dx_1 ~dx_2\eeq
in the range $0<x_1<1$ and $1-x_1 <x_2 <1$ where $x_i \equiv 2 q_i\cdot P/P^2$ such that $x_1+x_2+x_3 =2$.
The result is
\beq\label{eq:real}
\sigma v_{\rm rel} ({\rm DM}\,{\rm DM}\to q \bar q g)
= \sigma v_{\rm rel}|_{\rm tree} \left[ 1 + \frac{\alpha_3}{\pi} \left( \frac{4}{3\epsilon^2} +\frac{2-4\ell/3}{\epsilon}+\frac{19}{3}-\frac{7\pi^2}{9}-2\ell+\frac23\ell^2\right)\right]\eeq
where $\ell=\ln (4 M^2/\bar\mu^2)$ and
$ \sigma v_{\rm rel}|_{\rm tree}$ is the tree-level $ {\rm DM}\,{\rm DM}\to q \bar q$ cross section,
that depends on the DM and quark couplings (purely left-handed for multiplets with $Y=0$). 
Moving to the virtual correction, it arises solely from the vertex diagram in fig.\fig{MDMQCDannihilation}, as corrections to massless quark propagators vanish in dimensional regularization.
Doing the loop integration using Feynman parameters and integrating over them gives
\beq\label{eq:virtual}
\sigma v_{\rm rel} ({\rm DM}\,{\rm DM}\to q \bar q )= \sigma v_{\rm rel}|_{\rm tree} \left[ 1 - \frac{\alpha_3}{\pi} \left( \frac{4}{3\epsilon^2} +\frac{2-4\ell/3}{\epsilon}+\frac{16}{3}-\frac{7\pi^2}{9}-2\ell+\frac23\ell^2\right)\right].\eeq
After summing the real correction in eq.\eq{real} with the
virtual QCD correction in eq.\eq{virtual}, the infrared divergence cancels, yielding a finite result:
\beq 
\sigma v_{\rm rel} ({\rm DM}\,{\rm DM}\to q \bar q ) |_{\rm NLO} = \sigma v_{\rm rel}({\rm DM}\,{\rm DM}\to q \bar q )|_{\rm tree} \bigg[1 + \frac{\alpha_3}{\pi} \bigg]
\eeq
in agreement with analogous results from well-known processes such as $e^- e^+\to q \bar q$ and $Z\to q \bar q$.
The strong coupling $\alpha_3$ is here renormalized around $M$.

The total  $s$-wave annihilation cross section of
fermionic Minimal DM at tree level also includes final-state leptons, electroweak vectors and Higgs,
that do not receive QCD corrections.
The QCD correction to annihilations into quarks modifies the total $\sigma v_{\rm rel} $ into
\begin{equation}\label{eq:MDMtree}
\sigma v_{\rm rel} = 
\frac{g_2^4 (n^2-1)(2n^2+19+18\alpha_3/\pi) + 4  Y^2 g_Y^4 (41+8Y^2 + 22 \alpha_3/\pi)+16g_2^2 g_Y^2 Y^2 (n^2-1)}{256\pi\, M^2\, g_{\cal X}}
 \end{equation}
where an $n$-plet of $\SU(2)_L$ contains 
 $g_{\cal X}=2n$  degrees of freedom if it is a Majorana fermion with hyper-charge $Y=0$,
 or $g_{\cal X}=4n$ for a Dirac fermion (which is required when $Y \neq 0$, and also possible when $Y = 0$).
 The numerical coefficients in eq.\eq{MDMtree} account for the fact that three out of four fermion doublets in the Standard Model are quarks, and properly include their hypercharges,
 as well as annihilations into Higgses. 
Extra corrections from top Yukawa couplings are smaller, as they only affect  top quark and Higgs production.

\medskip

Eq.\eq{MDMtree} shows that QCD corrections are relatively more important at low $n$, as quark production becomes subdominant compared to purely electroweak channels at larger $n$.
For a fermionic {\em doublet} with $Y=1/2$ the QCD correction to the total $\sigma v_{\rm rel}$  is $1 + 0.63\alpha_3/\pi \approx 1.02$.
Since the DM relic abundance decreases approximately as
$\Omega_{\rm DM} \propto 1/\sigma$,
the DM mass required to reproduce the observed DM abundance grows as $M \propto \sqrt{\sigma}$, by about $+1\%$.

\medskip

For larger multiplets one needs to take into account that  the annihilation rates are significantly corrected by weak long-range interactions,
that multiply the cross sections by Sommerfeld factors $S_\lambda$ and induce extra bound-state processes.
In the $\SU(2)_L$-invariant limit these interactions are described by Coulombian potentials $V = - \lambda \alpha_2/r$ with $\lambda = (1+I^2-2n^2)/8$
for two-body DM states with iso-spin $I$.
DM annihilations into weak vectors take place when the initial DM DM state has total spin $S=0$ and $I = \{1,5\}$,
and are not affected by QCD corrections.
DM annihilations into fermions and into the SM Higgs doublets take place when the initial state has $S=1$ and $I=3$:
for $Y=0$ all such rates receive the QCD correction
\beq K_{\rm QCD}=1 + \frac{18}{25}\frac{\alpha_3}{\pi}.\eeq 
In particular, the Sommerfeld-corrected annihilation cross section of a fermion {\em triplet}  with $Y=0$ is
\begin{equation}
\sigma  v_{\rm rel} =\frac {37}{12}\, \frac {\pi \alpha_2^2}{M^2} \left[\frac {16}{111} S_2+\frac {75}{111}   K_{\rm QCD} S_1+ \frac {20}{111}  S_{-1}  \right].
\end{equation}
The total cross section receives a $1.3\%$ QCD enhancement in the $\SU(2)_L$-invariant approximation.
However, the $S=1$ rate gets  Boltzmann suppressed by $\SU(2)_L$-breaking effects at small temperature
$T \lesssim \Delta M \approx 166\MeV$~\cite{1702.01141}.
Furthermore, there is one bound state, with $S=\ell=0$: its annihilation rate into weak vectors receives no QCD correction at one loop,
and this bound state negligibly affects the relic abundance~\cite{1702.01141}.
Combining these effects, we find that the QCD correction decreases
the DM relic abundance  by $1.2\%$, and consequently increases the thermal DM mass  by $0.6\%$,
a shift comparable to the uncertainties estimated in~\cite{2107.09688} or~\cite{2410.02723}.


The Sommerfeld-corrected annihilation cross section of a fermionic {\em quintuplet} with $Y=0$ becomes, in the $\SU(2)_L$-invariant limit,
\begin{equation}
\sigma  v_{\rm rel} = \frac{207}{20}
 \frac {\pi \alpha_2^2}{M^2}\left[\frac {16}{69} S_6+ \frac {25}{69}K_{\rm QCD} S_5 +\frac {28}{69}  S_3\right]
\end{equation}
which amounts to a $+0.5\%$ QCD enhancement.
Furthermore, significant additional annihilation channels proceed via the formation of DM-DM bound states, 
 in which the excess energy is released through the emission of an electroweak gauge boson~\cite{1702.01141}.
The annihilation rates of bound states with $S=1$  into fermions and Higgses get enhanced by $K_{\rm QCD}$.
Such QCD corrections are relevant because, in the Boltzmann equations that dicate the relic abundance,
bound state annihilation rates compete with bound state breaking rates~\cite{1702.01141}.
Taking into account Sommerfeld corrections, $\SU(2)_L$ breaking and bound states as in~\cite{1702.01141},
we find that the QCD correction  leads to a $0.5\%$ reduction in the DM relic abundance $\Omega_{\rm DM}$.
Thereby  the quintuplet mass favoured by cosmology increases by $0.25\%$.

%

\medskip

In conclusion, we computed the QCD corrections to the annihilation rates of Minimal Dark Matter multiplets.
These corrections  reduce the thermal relic masses $M$ of the dark matter candidates at per-cent level.
We do not present a full updated computation of $M$ because 
additional loop-level effects of comparable size are expected and must be accounted for to reach consistent accuracy.
Indeed, weak corrections to DM annihilations contain terms enhanced by the multiplicity $n$ of the DM multiplet.
An example is the correction $\Delta b_2 = n(n^2-1)/18$ to the running of $\alpha_2$.
These sizeable weak corrections are only partially included through the Sommerfeld enhancement and bound-state formation processes. 
One expects additional corrections not enhanced by the small DM velocity, and thereby
of order $1 \pm \Delta b_2 \alpha_2/4\pi $~\cite{Harris:1957zza}, 
that can amount to a few $\%$ for $n=5$.



%

\footnotesize

\begingroup
\renewcommand{\baselinestretch}{0.95}\selectfont

\endgroup

\end{document}